# Knowledge Engineering Technique for Cluster Development

Pradorn Sureephong[1], Nopasit Chakpitak[1], Yacine Ouzroute[2], Gilles Neubert[2], and Abdelaziz Bouras[2]

[1] College of Arts, Media and Technology, Chiang Mai University. 239, Huawkaer Rd. T.Suthep A.Mueang 50200 Chiang Mai, Thailand
dorn@camt.info, nopasit@camt.info
[2] LIESP Laboratory, University Lumiere Lyon 2. 160 Boulevard de l'Université 69676 Bron Cedex, Lyon, France
firstname.lastname@univ-lyon2.fr

**Abstract.** After the concept of industry cluster was tangibly applied in many countries, SMEs trended to link to each other to maintain their competitiveness in the market. The major key success factors of the cluster are knowledge sharing and collaboration between partners. This knowledge is collected in form of tacit and explicit knowledge from experts and institutions within the cluster. The objective of this study is about enhancing the industry cluster with knowledge management by using knowledge engineering which is one of the most important method for managing knowledge. This work analyzed three well known knowledge engineering methods, i.e. MOKA, SPEDE and CommonKADS, and compare the capability to be implemented in the cluster context. Then, we selected one method and proposed the adapted methodology. At the end of this paper, we validated and demonstrated the proposed methodology with some primary result by using case study of handicraft cluster in Thailand.

## 1 Introduction

The knowledge-based economy is affected by the increasing use of information technologies. Most of industries try to use available information to gain competitive advantages[1]. From the study of ECOTEC in 2005[2] about the critical success factors in cluster development, first two critical success factors are *collaboration* in networking partnership and *knowledge creation* for innovative technology in the cluster which are about 78% and 74% of articles mentioned as success criteria accordingly. This knowledge is created through various forms of local inter-organizational collaborative interaction [3].

Study of Yoong and Molina [4] assumed that one way of surviving in today's turbulent business environment for business organizations is to form strategic alliances or mergers with other similar or complementary business companies. Thus, grouping as a cluster seems to be the best solution to increase the competitiveness for companies[5][6]. Although, many literatures claimed that knowledge is very important for cluster development but there is no empirical method to initiate or improve knowledge sharing for cluster.

Developing knowledge-based application creates difficulties to knowledge engineers[7]. Knowledge-based project cannot be handle by general software engineering methodology. The lifecycle of knowledge based application and software application is different in many aspects. In order to achieve the objective of knowledge engineering, Knowledge-Based Engineering (KBE) application lifecycle [8] focuses on these six critical phases as shown in figure 1.

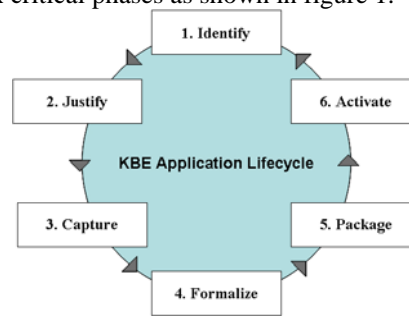

**Fig. 1.** Knowledge based engineering application lifecycle

### 1.1 Knowledge Engineering Techniques

Actually, knowledge is not a new idea [9] [10], philosophers and scholars had been studying it for centuries. There are many knowledge engineering techniques which are used for solving problem. However, we choose well known techniques that widely used in many projects for this study, i.e. MOKA, SPEDE and CommonKADS. These techniques were applied in different projects in various domains. All these methods are based on this KBE application lifecycle [8] which focuses on six critical phases as shown in fig. 1. We will analyze their capacity to be implemented in the cluster context.

**MOKA** (*M*ethodology and tools *O*riented to *K*nowledge based engineering *A*pplications) aims to help the structure side of the capture process (Fig. 1) in KBE application lifecycle. It focuses on two levels of representation - an informal and a formal model. These models provide the means of recording the structure behind the knowledge - including not only things about the product and design process but about the design rationale as well. Informal model is assembled from five categories of knowledge types; described on forms, known as ICARE forms (*I*llustrations, *C*onstraints, *A*ctivities, *R*ules and *E*ntities)[11].

**SPEDE** (*S*tructured *P*rocess *E*licitation *D*emonstrations *E*nvironment) provides an effective means to capture, validate and communicate vital knowledge to provide business benefit [12]. The concept of this methodology is to develop a guide that suitable for using with a variety of BPR/BPI projects, while at the same creating a means of enhancing process. This was made possible by breaking the activities into generally acknowledged high level BPR/BPI stages and providing a sequence at that level. The concept of this methodology is called *Swim Lane*. SPEDE provides Knowledge Acquisition (KA) tools to facilitate and assist the process in knowledge context.

**CommonKADS** (*Common K*nowledge *A*cquisition and *D*esign *S*ystem) is a methodology to support structured knowledge engineering. It provided *CommonKADS model suite* for creating requirements specification for knowledge system. The organization, task, and agent models analyze the organizational environment and the corresponding critical success factors for a knowledge system. The knowledge and communication models yield the conceptual description of problem-solving functions and data that were handled and delivered by a knowledge system. The design model converts it into a technical specification that is the basics for software system implementation [13].

### 1.2 Knowledge Engineering Technique Selection

In the study, we used knowledge-based engineering lifecycle and their provided tools as our criteria to select knowledge engineering technique for this study. The result of the comparison was shown in Table. 1.

**Table 1. Three methods compared with Knowledge Based Engineering Lifecycle**

| KBE Lifecycle | MOKA | SPEDE | CommonKADS |
|---|---|---|---|
| 1. Identify | - | Understand the project | Context Level |
| 2. Justify | - | Understand the project | OTA Model |
| 3. Capture | Informal Model | Design the process | Concept Level |
| 4. Formalize | Formal Model | Evaluate the new process | Concept Level |
| 5. Package | - | Communicate Process | Artifact Level |
| 6. Activate | - | - | - |

MOKA focuses on capturing and formalizing knowledge [11] to solve the specific engineering problems. On the other hand, we found that both SPEDE and CommonKADS techniques support KBE lifecycle from knowledge identifying to packaging phase. Then, we consider in the detail of each models of SPEDE and CommonKADS. SPEDE provided swim lane flowcharts as tools for each processes. SPEDE technique mainly focused on business process improvement in knowledge context. CommonKADS provided models and templates for each level. These templates support knowledge engineer for knowledge elicitation in different knowledge task, i.e. *analytic tasks* and *synthetic tasks*. These help knowledge engineer to be able to apply this technique in different type of knowledge problems. From two criteria, support KBE lifecycle and provided tools, CommonKADS technique is suitable for applying with industry cluster problems.

## 2 The Proposed Methodology

The concept of proposed methodology is adopted from CommonKADS methodology which was divided into three levels called *CommonKADS model suite*, i.e. Context Level, Concept Level and Artifact Level. However, managing structured knowledge in the industry cluster is different from single organization in many aspects because of characteristic of the organization. For example, there is no single policy maker in the cluster. So, KE could not utilize Context level's worksheets to assess single company for developing KMS for all companies within the cluster.

### 2.1 Context Level

This level contain organization, task and agent model. The main objectives of this level are giving the scope and clear view of the organization, knowledge intensive tasks and actors who involved in the task. It also provides the impact assessment, changes and consensus for knowledge engineering project. Knowledge Engineer (KE) should start at the most influence association in the cluster. Due to, association always be a group of potential companies in the industry and able to set policy/direction for the industry. KE could utilize Organization Mode (OM) worksheets for interviewing with associations. Then, the outputs from OM are knowledge intensive tasks from broken down process and agents who are related to each task. Then, KE could interview with experts in each task by using TM and AM worksheet. Finally, KE validate the result of each module with association again to assess impact and changes with OTA worksheet.

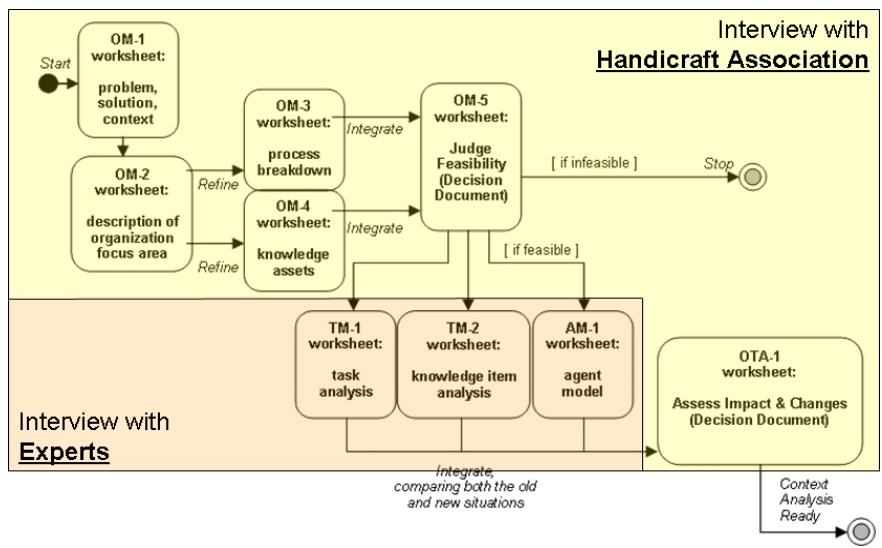

**Fig. 2.** A road map for carrying out knowledge oriented organization and task analysis.

### 2.2 Concept Level

The second level contain knowledge and communication model. This level is related with capturing and formalizing phases in KBE lifecycle. The main objectives of this to the level are to explicate in detail the types and structures used in performing task and to model the communicative transaction between the agents involved [13]. We acquired Task knowledge from context level which composed of task goal, decomposition and control. Inference and Domain knowledge could be obtained from knowledge elicitation process.

### 2.3 Artifact Level

The last level is the artifact level, contain design model. This level is related to packaging phase in KBE lifecycle or application development. The main objectives of this level are to give the technical system specification, such as architecture, implement platform, software modules, representation constructs, and computational mechanisms needed to implement the functions laid down in the knowledge and communication models [13].

## 3 Validation and Results

The initial investigations have been done with 10 firms within the two biggest handicraft associations in Thailand and Northern Thailand. *NO*rthern *H*andicraft *M*anufacturer and *EX*porter (NOHMEX) association is the biggest handicraft association in Thailand which includes 161 manufacturers and exporters. Another association which is the biggest handicraft association in Chiang Mai is named Chiang Mai Brand. It is a group of qualified manufacturers who have capability to export their products and passed requirements of Thailand's ministry of commerce. Until end of 2006, there are 99 authorized enterprises to use Chiang Mai brand on their products[14].

At the beginning of this study, CommonKADS was used as a knowledge engineering methodology in the context level (organization model, task model, and agent model) in order to understand organization environment and corresponding critical success factors for knowledge system.

As shown in Fig. 2, Organization Model (OM-1 to OM-5), we found that handicraft cluster has its own vision as *"Knowledge sharing hub for handicraft exporter"*. And, companies defined their problems, such as intellectual property problem, lack of collaboration, CDA development, product innovation, and product exporting. However, this cluster has many opportunities and solutions as well. We used "product exporting" and "product innovation" as our mock-up problems due to these problems is knowledge intensive and feasible in business and technical aspect.

From the Task Model (TM-1 to TM-2), we analyzed the feasibility of each tasks that related to product exporting and product innovation processes. This model makes it possible to rank and prioritize the different knowledge-improvement scenarios.

Agent Model (AM-1) proposed organizational recommendations, improvements, and actions. From the experts' point of view, they proposed actions for solving product exporting and product innovation problems as follow,

1) Develop information system that provides knowledge from experts about product selection, marketing information, or economic data from government organization.
2) Archive past lesson learn or experiences with in electronic forum
3) Create best practice of each task and store in knowledge-based system
4) Increasing the collaboration and information sharing within the cluster.
5) Create tools to support the capability of cluster development agency (CDA) to facilitate the cluster.